\documentclass{ws-procs975x65}

\begin{document}

\title{Induced Compression of White Dwarfs by Angular Momentum Loss}

\author{Kuantay Boshkayev,$^{1,2,*}$  Jorge A. Rueda,$^{2}$ Remo Ruffini$^{2}$ and  Bakytzhan Zhami.$^{1}$}

\address{$^1$Faculty of Physics and Technology, IETP, Al-Farabi Kazakh National University,\\
Al-Farabi avenue 71, Almaty, 050040, Kazakhstan\\
$^2$International Center for Relativistic Astrophysics Network,\\
Piazza della Repubblica 10, Pescara, I-65122, Italy\\
$^*$E-mail: kuantay@mail.ru}

\begin{abstract}

We investigate isolated sub- and super-Chandrasekhar white dwarfs which lose angular momentum through magnetic dipole braking. We construct constant rest mass sequences by fulfilling all stability criteria of rotating configurations and show how the main structure of white dwarfs such as the central density, mean radius and angular velocity change with time. We explicitly demonstrate that all isolated white dwarfs regardless of their masses, by angular momentum loss, shrink and increase their central density. We also analyze the effects of the structure parameters on the evolution timescale both in the case of constant magnetic field and constant magnetic flux.

\end{abstract}


\bodymatter
\
\section{Introduction}\label{sec:1}

In accordance with the current understanding, white dwarf (WD) stars represent the final product of the evolution of progenitor stars with the masses ranging from roughly 0.07 to 8 $M_{\odot}$ (solar mass). \cite{fontaine2001} Most of the observed WDs masses are clustered around (0.5-0.7)$M_{\odot}$,\cite{kepler1} whereas their size is typically to that of a planet. Hence they possess large average densities, huge surface gravities and low luminosities. \cite{shapiro1983}

In order to explain the formation of low mass (sub-Chandrasekhar) WDs it is believed that their progenitors were main sequence stars of low or medium mass.\cite{fontaine2001} However to explain the formation of massive white dwarfs (super-Chandrasekhar) there exist several scenarios: a single degenerate scenario,\cite{ whelan1973, nomoto1982, han2004} where a WD increases its mass through accretion of matter from a normal stellar companion; a double degenerate scenario,\cite{iben1984, webbink1984, kerkwijk2010} where two average mass WDs merge after losing energy and angular momentum through the radiation of gravitational waves; and a core degenerate scenario,\cite{ kashi2011} where the merger occurs in a common envelop with a massive asymptotic branch star.\cite{ilkov2012}

In this regard we investigated general relativistic configurations of uniformly rotating sub- and super-Chandrasekhar WDs in our recent work\cite{2013ApJ...762..117B} within Hartle's formalism.\cite{1967ApJ...150.1005H, HT1968} We used the relativistic Feynman-Metropolis-Teller (RFMT) equation of state\cite{2011PhRvC..83d5805R} (EoS) for WD matter, which generalizes the traditionally used equations of state of Chandrasekhar\cite{chandrasekhar31} and Salpeter.\cite{1961ApJ...134..669S} The stability of rotating WDs was analyzed taking into account the mass-shedding limit, inverse $\beta$-decay instability, pynonuclear instability, and secular axisymmetric instability, with the last being determined by using the turning point method of Friedman et al.\cite{1988ApJ...325..722F} 

In this work, we explore the compression of isolated rotating sub- and super-Chandrasekhar WDs by angular momentum loss based on the results of Ref.~\refcite{2013ApJ...762..117B}. Eventually, by fulfilling all the stability criteria for rotating WDs in general relativity (GR) we calculate the change in time of the basic parameters of rotating WDs. We consider two particular cases: a) we assume that dipolar magnetic is constant throughout the entire evolution of WDs; b) we adopt the magnetic flux conservation, relaxing the constancy of the magnetic field, for the sake of comparison. 
%
%%%%%%%%%%%%%%%%%%%%%%%%%%%%%%%%%%%%%%%%%%%%%%%%%%%%%%%%%%%%%%%%%%%%%%%%%%%%%%%%%%%%%%%%%%%%%%%%%%%%%%%%%%%%%%%%%%%%
%%%%%%%%%%%%%%%%%%%%%%%%%%%%%%%%%%%%%%%%%%%%%%%%%%%%%%%%%%%%%%%%%%%%%%%%%%%%%%%%%%%%%%%%%%%%%%%%%%%%%%%%%%%%%%%%%%%%
\section{Compression of isolated rotating white dwarfs}
\label{sec:4}

In order to examine the compression of isolated white dwarfs with time, first, taking into account all the stability criteria, we constructed constant mass sequences for selected sub- and super-Chandrasekhar WDs. Second, knowing how the main parameters change along the sequences we find their dependence on each other. Third, using the equation for the rotational energy loss of pulsars via magnetic dipole braking,\cite{malheiro2012} we estimate the evolution time that is required to go from one instability point to another.\cite{boshizzo} Fourth, depending on what parameter we are interested in, we modify the equation and analyze how this parameter changes in time.

For example, if we want to examine how the central density evolves with time we need to rewrite the equation as follows
\begin{equation}\label{eq:rhot}
dt=-\frac{3}{2}\frac{c^3}{ B^2}\frac{1}{{\langle R \rangle}^6}\frac{1}{\Omega^3}\frac{\partial J}{\partial \rho} d\rho,
\end{equation}
where we choose all parameters as a function of the central density $\rho$
\begin{equation}
\langle R\rangle=\langle R\rangle(\rho),\qquad \Omega=\Omega(\rho), \qquad J=J(\rho)
\end{equation}
and $t$ is the evolution time, $c$ is the speed of light in vacuum, $B$ is the dipole magnetic field, $\langle R\rangle=(1/3)(R_p+2R_e)$ is the mean spherical radius, $R_p$ is the polar radius, $R_e$ is the equatorial radius, $\Omega$ is the angular velocity and $J$ is the angular momentum of the white dwarf. We adopt the maximum brake, so that the angle between the rotation axis and the magnetic field is ninety degrees. The values of $\langle R\rangle=\langle R \rangle(\rho),\quad\Omega=\Omega(\rho)$ and  $J=J(\rho)$ are calculated along a given constant baryon (rest) mass sequence as mentioned above. Here we choose carbon white dwarfs using the RFMT EoS as in Ref.~\refcite{RotD2011}.

\begin{figure}[t]
\centering{\includegraphics[width=0.75\columnwidth,clip]{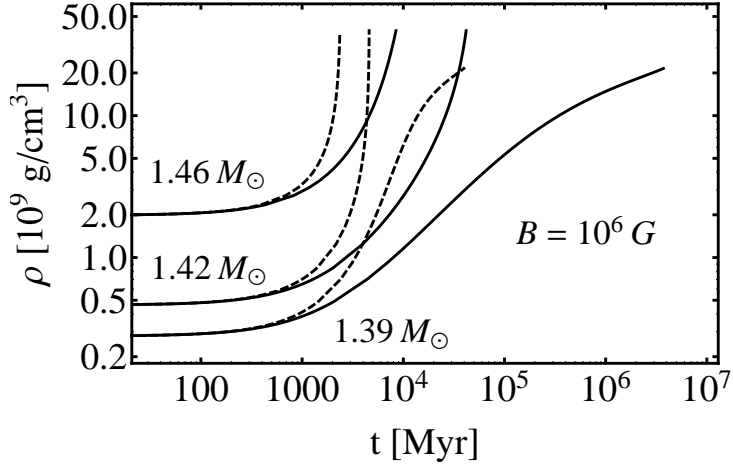}} 
\caption{Central density versus time. Solid curves are the evolution paths for selected constant mass sequences when $B$ is constant. Dashed curves are the evolution paths for constant mass sequences when the magnetic flux is constant for $B_0=10^6$ G.}\label{fig:rhoti}
\end{figure}

According to Ref.~\refcite{2013ApJ...762..117B} and \refcite{boshizzo,boshijmpe,boshijmpcs,boshjkps,boshmg13} there are limiting values of all the parameters of rotating WDs on the border of the stability region. These limiting values determine the range of integration of the central density $\rho$. Performing numerical integration of Eq.~\ref{eq:rhot} for each moment of time we obtain the evolution of the central density with time. We use the following expression to account for the magnetic flux conservation, relaxing the constancy of the magnetic field 

\begin{equation}\label{eq:flux}
B=B_0\frac{{\langle R_0\rangle}^2}{{\langle R\rangle}^2},
\end{equation}
where $B_0$ is the surface dipole magnetic field corresponding to the initial value of $B$ at $t=0$, ${\langle R_0\rangle}$ is the mean radius corresponding to the initial values of $\langle R\rangle$ at $t=0$. Substituting $B$ in Eq. \ref{eq:rhot} we obtain an equation that describes the evolution of $\rho$ with time accounting for the magnetic flux conservation. Clearly, in Fig.~\ref{fig:rhoti} we see the main difference between two cases. In both cases WDs will increase their central density and will be compressed by losing angular momentum through magnetic dipole braking. However in the case with the magnetic flux conservation the value of $B$ will be increasing due to the compression thus causing more torque and evolving faster with respect to the $B$=constant case.

\begin{figure}
\centering{\includegraphics[width=0.75\columnwidth,clip]{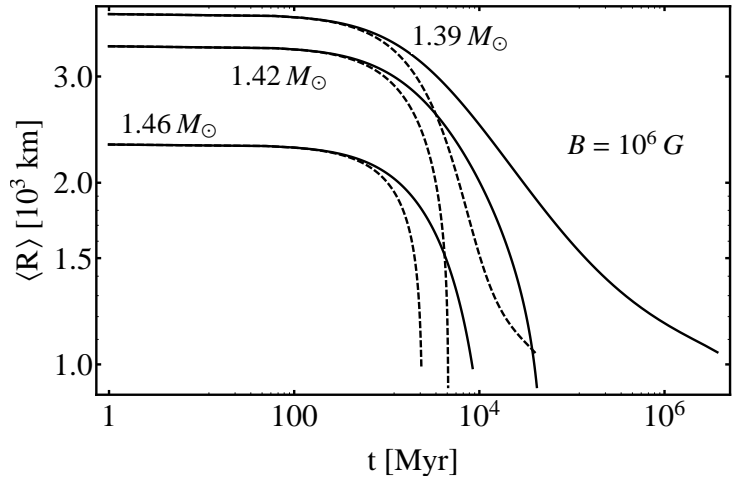} }
\caption{Mean radius versus time. Solid curves are the evolution paths for selected constant mass sequences when $B$ is constant. Dashed curves are the evolution paths for constant mass sequences when the magnetic flux is constant for $B_0=10^6$ G.}\label{fig:rmeanti}
\end{figure}

\begin{figure}
\centering{\includegraphics[width=0.75\columnwidth,clip]{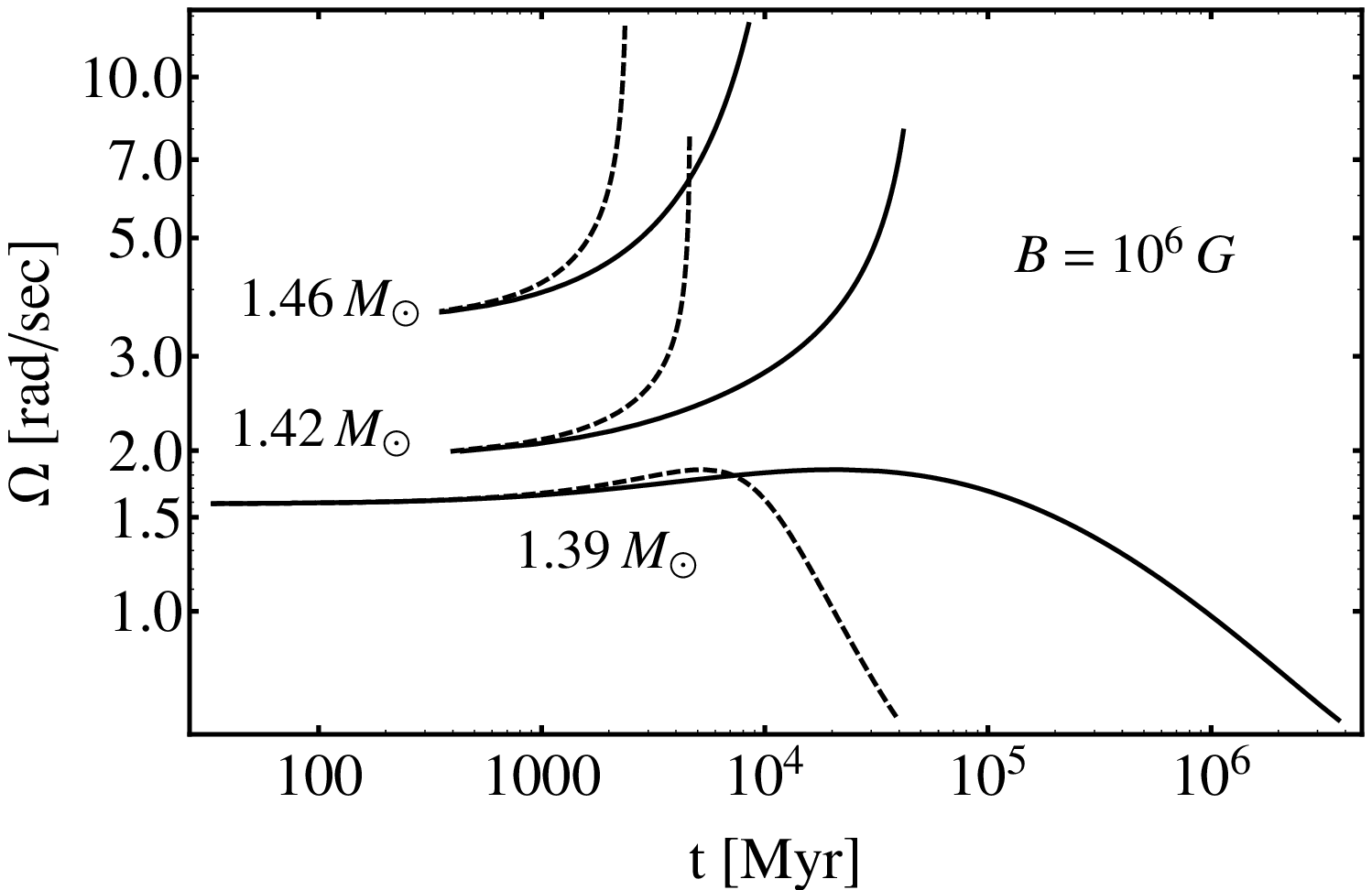}}
\caption{Angular velocity versus time. Solid curves are the evolution paths for selected constant mass sequences when $B$ is constant. Dashed curves are the evolution paths for constant mass sequences when the magnetic flux is constant for $B_0=10^6$ G.}\label{fig:oti}
\end{figure}

To calculate how the mean radius of the WD evolves with time we need to rewrite Eq.~\ref{eq:rhot} as follows

\begin{equation}\label{eq:Rmeant}
dt=-\frac{3}{2}\frac{c^3}{B^2}\frac{1}{{\langle R\rangle}^6}\frac{1}{\Omega^3}\frac{\partial J}{\partial\langle R\rangle} d \langle R \rangle,
\end{equation}
where now
\begin{equation}
\Omega=\Omega(\langle R\rangle), \qquad J=J(\langle R\rangle)
\end{equation}

The values of $\Omega=\Omega(\langle R\rangle)$ and  $J=J(\langle R\rangle)$ are calculated along a given constant rest mass sequence like in the previous case. Here the range of integration of $\langle R\rangle$ is also defined along the constant rest mass sequence on the border of the stability region. In Fig.~\ref{fig:rmeanti} the mean (average) radius is plotted as a function of time. Over the course of time the mean radius decreases, hence WDs shrink with time. For the case of the conserved magnetic flux the mean radius decreases faster than for the case with constant surface magnetic field. 

The procedure to estimate the evolution of other parameters of rotating WDs is analogous to the central density with Eq.~\ref{eq:rhot} and to the mean radius with Eq.~\ref{eq:Rmeant}. To take into consideration the magnetic flux conservation we need to use Eq.~\ref{eq:flux}. Consequently, in Fig.~\ref{fig:oti} we plot angular velocity of WDs as a function of time. Here we see that a WD of mass $1.39 M_{\odot}$ spins-up at the beginning and spins-down at the end, whereas super-Chandrasekhar WDs spin-up only. This is another confirmation of our previous calculations (see e.g. Refs.~\refcite{2013ApJ...762..117B,boshjkps}).

Isolated WDs regardless of their masses will always lose angular momentum via magnetic dipole braking. By losing angular momentum WDs tend to get more stable configuration by increasing their central density and by decreasing their mean radius. Eventually super-Chandrasekhar WDs will spin-up and sub-Chandrasekhar WDs spin-down. However WDs having masses close to the Chandrasekhar mass limit at certain time of their evolution will experience both spin up and spin down epochs. Fig.~\ref{fig:oti} is another illustration of the spin-up and spin-down epochs.\cite{boshjkps}

\section{Conclusion}
\label{sec:5}
%%%%%%%%%%%%%%%%%%%%%%%%%%%%%%%%%%%%%%%%%%%%%%%%%%%%%%%%%%%%%%%%%%
In this work, in view of our previous computations\cite{2013ApJ...762..117B} we adopted that we have various stable rotating WDs at different times in their evolutionary track without involving the details and the routes of their entire evolution. In this regard, we calculated constant rest mass sequences for selected masses and constructed tables showing the dependence of the WD's parameters on each other along the sequences. 

Furthermore, by means of the equation for the rotational energy loss of pulsars through magnetic dipole braking, we showed how the main parameters of rotating WDs evolve with time. For the sake of comparison we considered two cases with constant magnetic field and constant magnetic flux. For the magnetic flux conservation case the evolution time (life time) turned out to be shorter than for the constant magnetic field. Consequently, we showed that all uniformly rotating WDs will be compressed by angular momentum loss via magnetic dipole braking in order to pass from one equilibrium state to another.

We performed all computations in GR by using the Hartle formalism for uniformly-rotating configurations for the sake of generality. By exploiting the relativistic Feynman-Metropolis-Teller equation of state, we considered mainly WDs consisting of carbon, although the typical white dwarfs are known not to consist of a pure chemical element, but a mixture of elements such as carbon, oxygen, neon, magnesium, etc. The consideration of different chemical compositions will be the issue of our future investigations.
\section*{Acknowledgments}
This work was supported by Grant No. 3101/GF4 IPC-11 of the Ministry of Education and Science of the Republic of Kazakhstan. K.B. acknowledges ICRANet for its hospitality and support. %The authors thank the anonymous referee for useful recommendations, comments and suggestions which helped to improve the paper.  

\end{document}